Global ionospheric response to a periodic sequence of HSS/CIR events during the 2007-2008 solar minimum

Negrea C.[1], Munteanu C.[2], Echim M. M.[1,2]

[1] Institute of Space Science, Magurele, Romania

[2] Belgian Institute for Space Aeronomy, Brussels, Belgium

Corresponding author: Cătălin Negrea (negreacatalin@spacescience.ro)

Key points:

- The global ionospheric low-frequency response to a sequence of recurrent structures in the solar wind is analyzed using ionosonde data.

- The ionospheric $NmF_2$ and $hmF_2$ spectra indicate that nonlinear interactions transfer power from the main spectral peaks to secondary ones.

- The ionospheric spectra correlate best with the solar wind magnetic field $B_z$ and geomagnetic index SYM-H.

Abstract

In this study, we investigate the global ionospheric impact of high-speed solar wind streams/corotating interaction regions (HSS/CIR). A series of ten such events are identified between December 1$^{st}$ 2007 and April 29$^{th}$ 2008, characterized in the frequency domain by the main spectral peaks corresponding to 27, 13.5, 9 and 6.75 days. The spectra of solar wind magnetic field, speed and proton density, as well as those of the geomagnetic indices AE and SYM-H are solely dominated by these features. By contrast, the ionospheric $NmF_2$ and to a lesser extent the $hmF_2$ spectra have a much more complex structure, with secondary peaks adding to or replacing the main ones. We argue that this is evidence of the nonlinear nature of the magnetosphere-ionosphere coupling, highlighted particularly in the $NmF_2$ ionospheric response. Additionally, we show that $hmF_2$ is more closely correlated than $NmF_2$ to all parameters describing the solar wind and geomagnetic activity. Finally, the ionospheric response shows higher correlation with $B_z$ than any other solar wind parameter, and higher with SYM-H than AE, indicating that for the low-frequency part of the spectrum, high-latitude Joule heating and particle precipitation play a secondary role to that of prompt penetration electric fields in dictating the ionospheric response to geomagnetic activity, in the case of this sequence of HSS/CIR events.

1. Introduction

The thermosphere-ionosphere system constitutes the upper-most layer of the Earth's atmosphere and is characterized by high variability and a strong dependence on external drivers. The two main categories of forcings are solar radiation and geomagnetic activity at high latitudes. Under normal conditions, solar radiation is the dominant source of energy and momentum, generating the standard structure, composition and circulation of the thermosphere-ionosphere (*Rishbeth and Garriott*, 1969; *Fuller-Rowell*, 2014). During geomagnetic storms, the energy input at high latitudes can be significant, or even dominant. During such events, there is an increase of particle precipitation and Joule heating (*Codrescu et al.*, 1995) at high latitudes, together with penetration electric fields at mid- and low-latitudes. The resulting plasma drift and associated ion-drag will heat up the neutral atmosphere, inducing pressure gradients that can alter the global neutral circulation from a dayside-nightside pattern of winds to an equatorward pattern. Additionally, since the Joule heating maximum occurs around 100-120 km altitude, the heavier molecular species dominant here (molecular oxygen and molecular nitrogen) will be displaced toward higher altitudes, while the atomic oxygen density will decrease due to equatorward winds (*Fuller-Rowell et al.*, 1994). The resulting modified chemical composition will in turn lead to a higher plasma loss rate through charge exchange and recombination, as well as a decrease in plasma production due to the loss of atomic oxygen.

The occurrence of geomagnetic storms is linked to solar activity and transient events in the solar wind. During the descending and minimum phases of a solar cycle, the near-Earth solar wind is dominated by high-speed streams (HSSs) and their associated stream interaction regions (SIRs), characterized by increased plasma density and magnetic field intensity. These structures are formed when the fast-solar wind emanating from a solar coronal hole (CH) "collides" with the slower solar wind ahead of it. Coronal holes lasting for several solar rotations will generate recurring SIRs, also known as corotating interaction regions (CIRs). The ionospheric impact of geomagnetic storms and of HSS/CIR sequences is an active area of research, with multiple recent studies employing a variety of data types and approaches: TEC (*Verkhoglyadova et al*, 2011, 2013; *Chen et al*, 2015), ionosonde data, primarily $NmF_2$ (*Grandin et al.*, 2015; *Wang et al*, 2011), satellite measurements from CHAMP (*McGranaghan et al.*, 2014; *Xu et al.*, 2015), TIMED (*Crowley et al.*, 2008), COSMIC (*Tulasi Ram et al.*, 2010a, 2010b), as well as modelling tools (*Klenzing et al.*, 2013, *Fuller-Rowell et al*, 1994; *Wang et al.*, 2011).

The early minimum phase of solar cycle 23 (2007-2008) was particularly dominated by the periodic recurrence of HSS/CIR events (*Guo et al.*, 2012; *Lei et al.*, 2011; *Tulasi Ram et al.*, 2010a). Munteanu et al. (2019) demonstrated the existence of two dominant sequences of HSS/CIR events recurring with similar properties over five solar rotations, from December 2007 to April 2008. They showed that each HSS/CIR event produced strong high-latitude activity (substorms) as observed in the AE index and a minor storm, as observed in the SYM-H index. The whole heliosphere interval (WHI) between March 20 and April 16, 2008 (Carrington rotation 2068) was part of an international campaign to study the solar-heliospheric-planetary system near the solar minimum between cycles 23 and 24 (*Echer et al.*, 2011; *Gibson et al.*, 2011; 2009; *Maris & Maris*, 2009; *Thompson et al.*, 2011; *Webb et al.*, 2011). *Lei et al.* (2011) showed that the HSSs/CIRs around the WHI interval were geoeffective both in terms of geomagnetic activity as well as thermospheric density response, with relative changes in neutral density due to HSSs/CIRs as large as 75%, which indicates that HSSs/CIRs have a significant impact on the variability of the thermosphere, even though the resultant geomagnetic activity is weak or moderate.

The recurring nature of the solar wind perturbations associated with HSS/CIRs induce recurring geomagnetic events characterized by periodicities of 27 days, as well as the higher harmonics at 13.5, 9 and 6.75 days (*Lei et al.*, 2011; *Katsavrias et al.*, 2012; *Crowley et al.*, 2008). Finally, the response of the thermosphere-ionosphere to these storms is characterized by the same periodicities (*Lei et al.*, 2008, 2011; *Pedatella et al.*, 2010; *Tulasi Ram et al.*, 2010a, 2010b; *Xu et al.*, 2015; Emery et al., 2011), as well as additional spectral peaks (*Wang et al.*, 2011).

In this study, we aim to provide a description of the global response of the ionosphere to the series of HSS/CIRs between December 2007 and April 2008. For this purpose, we used ionosonde $NmF_2$ and $hmF_2$ data from the NOAA-NCEI database (http://ngdc.noaa.gov/ionosonde/data), as well as complementary OMNI datasets (*King and Papitashvili*, 2005) for solar wind parameters and geomagnetic indices. We applied a spectral analysis method based on the Lomb-Scargle method to highlight dominant spectral features, while simultaneously taking into account the differences in sampling between the different datasets. To further account for differences caused by the geographical location and local conditions at each station, the spectra were individually normalized to 1, allowing for comparisons between results obtained at different locations. The results provide a comprehensive image of the global ionospheric response to the series of HSS/CIR events recorded in 2007-2008.

The paper is structured as follows: section 1 contains the introduction, section 2 describes the ionosonde $NmF_2$ and $hmF_2$ data, the OMNI solar wind measurements and geomagnetic indices, as well as the data processing and analysis methodology. Section 3 contains the main geophysical results and their discussion, and finally section 4 lists our conclusions.

2. Dataset and analysis methodology

The ionospheric response is studied using ionosonde $NmF_2$ and $hmF_2$ data publicly available through the National Oceanic and Atmospheric Administration – National Centers for Environmental Information (NOAA – NCEI) website. A total of 28 stations provided measurements for the period of interest (1 December 2007 – 30 April 2008), a complete list of which is given in table 1. At all locations, a threshold of 2 was used as a filter for the quality figure automatically generated by QualScan (McNamara, 2006), excluding poor quality data. Additionally, subgroups of stations are highlighted using a color code: dark-blue for high-latitude stations, dark-orange for mid-latitude northern hemisphere stations, dark-red for low-latitude locations and light-orange for the mid-latitude southern hemisphere. The data acquisition is performed independently at each location, and subject to different local conditions, such as magnetic latitude or weather patterns. This can potentially give rise to discrepancies in the dynamic range and sampling characterizing each station. A sample time series spanning four months of data from the RAL ionosonde is displayed in figure 1, and it shows the impact of 10 geomagnetic storms caused by recurring CIR/HSS events, where the areas shaded with red and blue denote the time periods associated with the two HSS/CIR sequences identified by *Munteanu et al.* (2019). Also, a 10-day long subset from the same location highlights the ionospheric response to two particular geomagnetic events, started around 26-03-2008 and 04-04-2008 in fig. 1b and fig. 1d. The typical storm ionospheric response is seen, characterized by a sudden increase in density at the outset of the storm (the positive phase), followed by a decrease in density lasting several days (the negative phase). It should however be noted that the specific ionospheric response to geomagnetic events can be more complex, particularly for moderate events (*Matamba et al.*, 2015), such as the ones investigated in this study.

*Table 1. List of ionosonde stations, together with geographic latitude and latitude, start and end-date for the available data, typical sampling frequency and percentage data coverage.*

| Station Code | Latitude (°) | Longitude (°) | Start Date | End Date | Sampling Frequency (min$^{-1}$) | Data Coverage (%) |
|---|---|---|---|---|---|---|
| THJ77 | 77.5 | 290.8 | 01.12.2007 | 29.04.2008 | 15 \| 5 | 73.7% |
| TR169 | 69.6 | 19.2 | 17.12.2007 | 29.04.2008 | 15 | 20.8% |
| SMJ67 | 66.98 | 309.06 | 01.12.2007 | 29.04.2008 | 15 \| 5 | 31.3% |
| CO764 | 64.9 | 212 | 01.12.2007 | 29.04.2008 | 15 | 36.7% |
| GA762 | 62.38 | 215 | 01.12.2007 | 29.04.2008 | 15 | 15.6% |
| NQJ61 | 61.2 | 314.6 | 01.12.2007 | 29.04.2008 | 15 | 28.6% |
| KS759 | 58.4 | 203.6 | 01.12.2007 | 29.04.2008 | 15 | 47.2% |
| JR055 | 54.6 | 13.4 | 01.12.2007 | 29.04.2008 | 15 | 48.0% |
| GSJ53 | 53.3 | 299.7 | 01.12.2007 | 29.04.2008 | 15 | 25.6% |
| FF051 | 51.7 | 358.2 | 01.12.2007 | 29.04.2008 | 15 | 79.1% |
| RL052 | 51.6 | 358.7 | 25.12.2007 | 29.04.2008 | 10 | 62.7% |
| PQ052 | 50 | 14.5 | 01.12.2007 | 29.04.2008 | 15 | 62.3% |
| MHJ45 | 42.6 | 288.5 | 01.12.2007 | 29.04.2008 | 7.5 \| 15 | 20.1% |
| RO041 | 41.9 | 12.5 | 01.12.2007 | 29.04.2008 | 15 | 58.2% |
| EB040 | 40.8 | 0.5 | 01.12.2007 | 29.04.2008 | 15 | 73.7% |
| BC840 | 40 | 255 | 01.12.2007 | 29.04.2008 | 15 | 60.4% |
| AT138 | 38 | 23.5 | 01.12.2007 | 29.04.2008 | 15 | 48.2% |
| WP937 | 37.9 | 284.5 | 01.12.2007 | 07.01.2008 | 15 | 20.2% |
| EA036 | 37.1 | 353.3 | 01.12.2007 | 29.04.2008 | 15 | 74.8% |
| PA836 | 34.7 | 239.4 | 01.12.2007 | 29.04.2008 | 15 | 64.2% |
| DS932 | 32.4 | 260.2 | 01.12.2007 | 29.04.2008 | 15 | 56.1% |
| KJ609 | 9 | 167.2 | 01.12.2007 | 29.04.2008 | 15 | 39.2% |
| AS00Q | -7.95 | 345.6 | 01.12.2007 | 29.04.2008 | 15 | 49.9% |
| JI91J | -12 | 283.2 | 01.12.2007 | 29.04.2008 | 15 | 61.1% |
| MU12K | -22.4 | 30.9 | 25.12.2007 | 29.04.2008 | 30 | 41.7% |
| LV12P | -28.5 | 21.2 | 25.12.2007 | 29.04.2008 | 30 | 47.8% |
| GR13L | -33.3 | 26.5 | 01.12.2007 | 29.04.2008 | 15 | 73.4% |
| PSJ5J | -51.6 | 302.1 | 26.12.2007 | 08.04.2008 | 30 | 11.3% |

In order to highlight the wide range of the natural ionospheric variability, as well as that of the response to geomagnetic storms, figure 2 illustrates the data at Millstone Hill. The extent of the differences in seasonal variability can be glean by comparing panels 1a,c with 2a,c, and the differences in ionospheric response are evident by comparing panels 1b,d with 2b,d. For example: the beginning of the negative storm phase at Millstone Hill is a full day later than at RAL for the event commencing on 25-03-2008, while for the event commencing on 04-04-2008, the relative change in $NmF_2$ is larger during the positive storm phase at Millstone Hill, and larger at RAL during the negative phase.

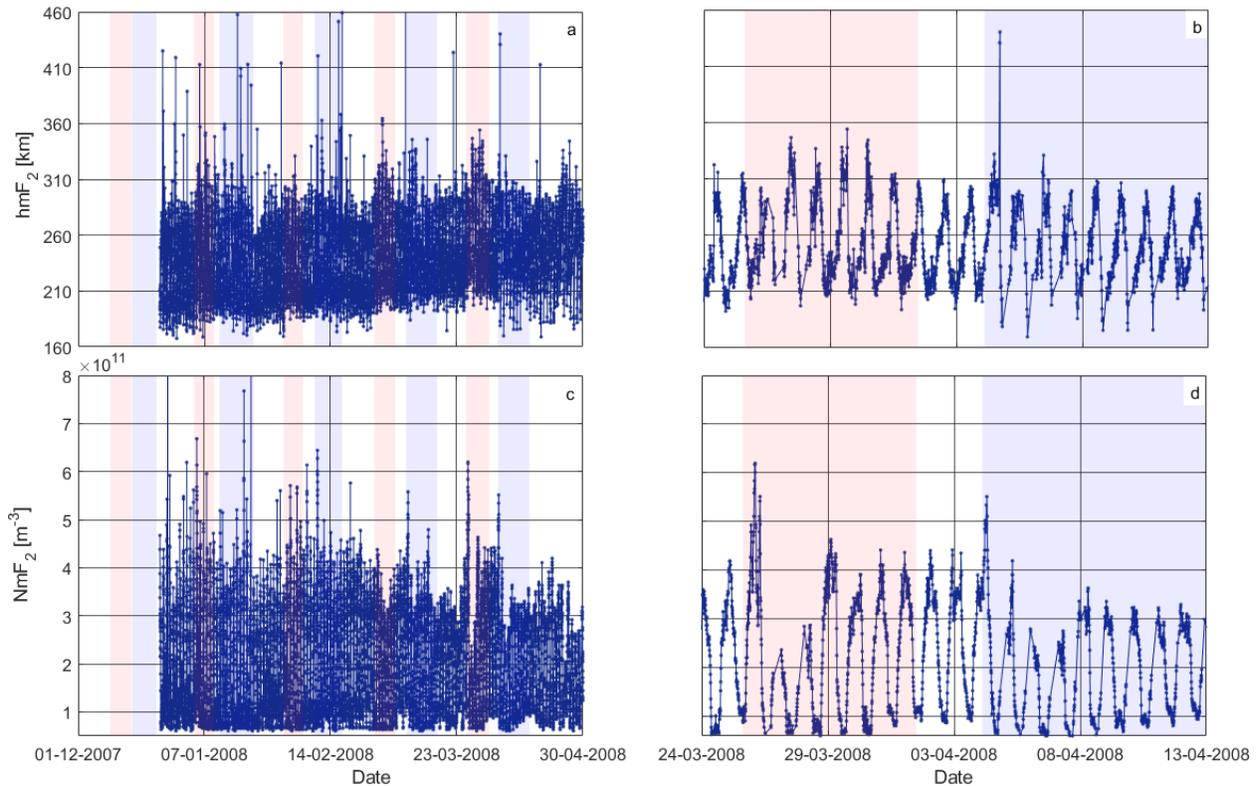

*Figure 1. Sample dataset from the RAL ionosonde (51.6°latitude, 358.7° longitude): a. full $hmF_2$ dataset, b. 20-day long $hmF_2$ subsets showing the impact of two consecutive CIR events, c. full*

$NmF_2$ *dataset and d. same as panel c, but for NmF2. The shaded areas deliniate time periods associated with HSS/CIR events.*

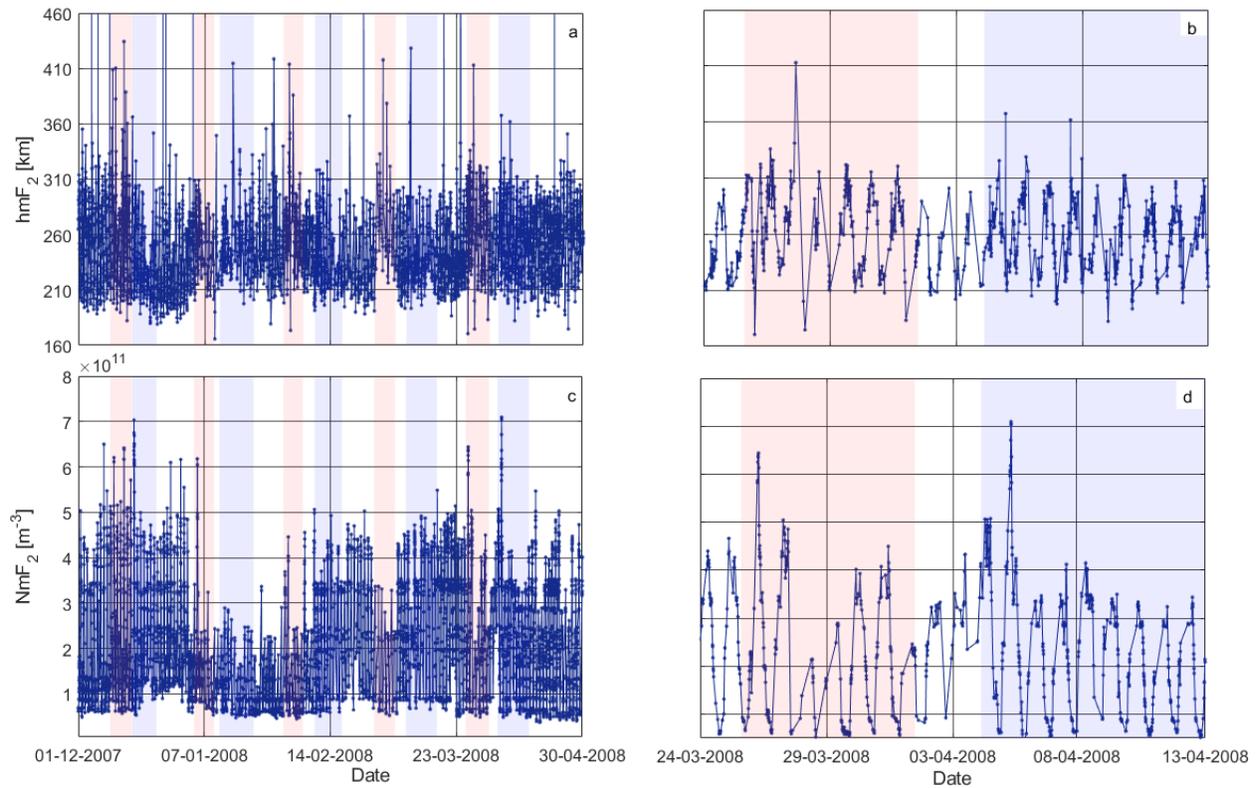

*Figure 2. Same as figure 1, but for the Millstone Hill (42.6° latitude, 288.5° longitude) ionosonde.*

The variations in solar wind velocity, density, magnetic field exhibit several periodicities linked to the interaction of these structures with the Earth's magnetosphere and are subject of several studies (see, e.g., *Munteanu et al.*, 2019). The magnetosphere itself reacts through a series of geomagnetic storms that follows in general the periodic trends of the interplanetary driving. However, some notable differences are observed as discussed by *Munteanu et al.* (2019). They demonstrated that the Russell-McPherron (R-M) effect (*Russell and McPherron*, 1973) systematically increased (decreased) the geoeffectiveness of the negative (positive) polarity

HSS/CIR streams. The R-M effect did not significantly affect the intensity of the storms caused by the R-M geoeffective streams, but rather increased the duration of their main phases with up to 3 days. Thus, the ionosphere is subject to two types of forcing, (i) the interplanetary forcing characterized by "pure" periodicities exhibited by solar wind parameters, and (ii) the geomagnetic/magnetospheric forcing characterized by recurrent/periodic occurrence of storms and substorms. A quantitative measure of the intensity of the magnetospheric forcing is given by geomagnetic indices like the auroral AE and equatorial SYM-H. A full set of parameters of interest is depicted in figure 3.

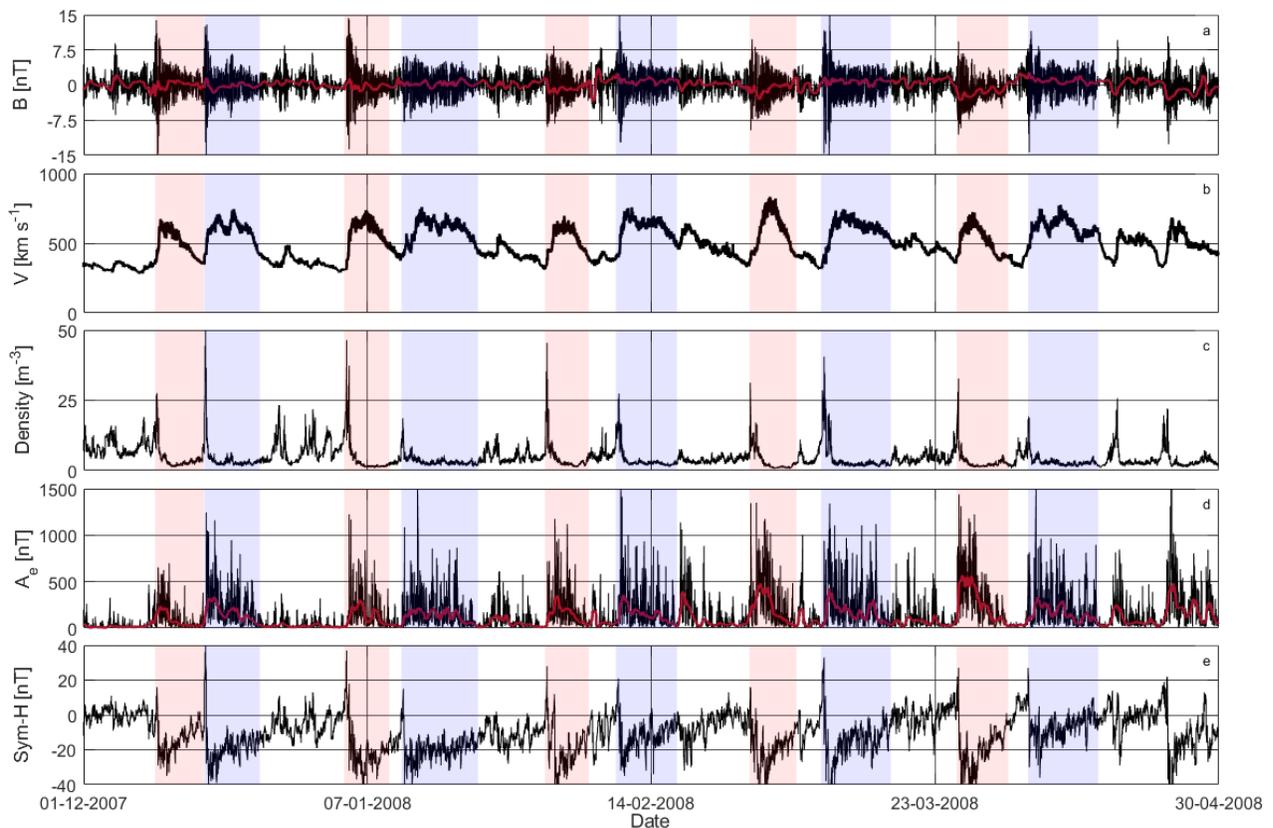

*Figure 3. Solar wind and geomagnetic indices for the same time period as that shown in fig. 1: a. solar wind magnetic field $B_z$ component; b. solar wind speed along the x-axis; c. AE index and d. SYM-H index.*

The sampling of the ionosonde data is generally uniform for 25 stations, while at 3 locations it varies between two values. Also, the sampling rates can differ from station to station, as can be seen in figures 1 and 2: at RAL the sampling rate is 1/10 min, while at Millstone Hill it oscillates between 1/7.5 min and 1/15 min. Additionally, this is different from the sampling rate of the solar wind measurements, which is 1/min. All datasets, with the exception of the geomagnetic indices, contain data gaps of varying sizes and distributions. Table 1 lists for each station, the start and end date for which measurements were available, as well as the overall data coverage and sampling frequency. The percentages are smaller than those for the solar wind measurements, which are 95.35% for the magnetic field $B_z$ component, 83.93% for solar wind speed, V and 83.92% for the proton density ($n_p$).

In order to account for these differences and obtain comparable spectra, a Lomb-Scargle implementation is used (*Lomb*, 1976; *Scargle*, 1982), allowing for equivalent spectra to be obtained for all locations and measurement types, despite the different amounts of data available and the presence of data gaps (*Munteanu et al.*, 2016). The amplitude spectra as a function of period is defined as:

$$X(P) = \sqrt{\frac{2}{N}\left[\frac{(\sum_i(x_i-\bar{x})\cos\omega(t_i-\tau))^2}{\sum_i \cos^2\omega(t_i-\tau)} + \frac{(\sum_i(x_i-\bar{x})\sin\omega(t_i-\tau))^2}{\sum_i \sin^2\omega(t_i-\tau)}\right]} \quad (1)$$

with the parameters: $\bar{x} = \frac{1}{N}\sum_i x_i$, $\tan(2\omega\tau) = \frac{\sum_i \sin 2\omega t_i}{\sum_i \cos 2\omega t_i}$, $\omega = \frac{2\pi}{P}$ is the angular frequency, P is the period and $N$ is the size of the dataset $(t_i, x_i)$. Figure 4 shows sample hmF$_2$ and NmF$_2$ spectra from the RAL ionosonde, covering the entire dataset. For periods of a day or more, the extent of the data gaps is unlikely to affect our results. The impact of several sources of variability can be seen: at periods of several hours or less, the spectrum reflects the impact of small-scale structures, predominantly gravity-wave (GW) induced travelling ionospheric disturbances (TIDs). Some of these disturbances are likely connected to geomagnetic activity, being either large scale TIDs (*Borries et al.*, 2009) or part of the background level of ionospheric disturbances (*Negrea et al.*, 2016, 2018). Superimposed are several dominant spectral peaks corresponding to the diurnal variation and its harmonics (periods of 24, 12, 8, 6, etc hours). While both these are impacted by geomagnetic activity, the main ionospheric response is located at low frequency, where spectral peaks are observed with periods of approximately $P_1=27$, $P_2=13.5$, $P_3=9$ and $P_4=6.75$ days.

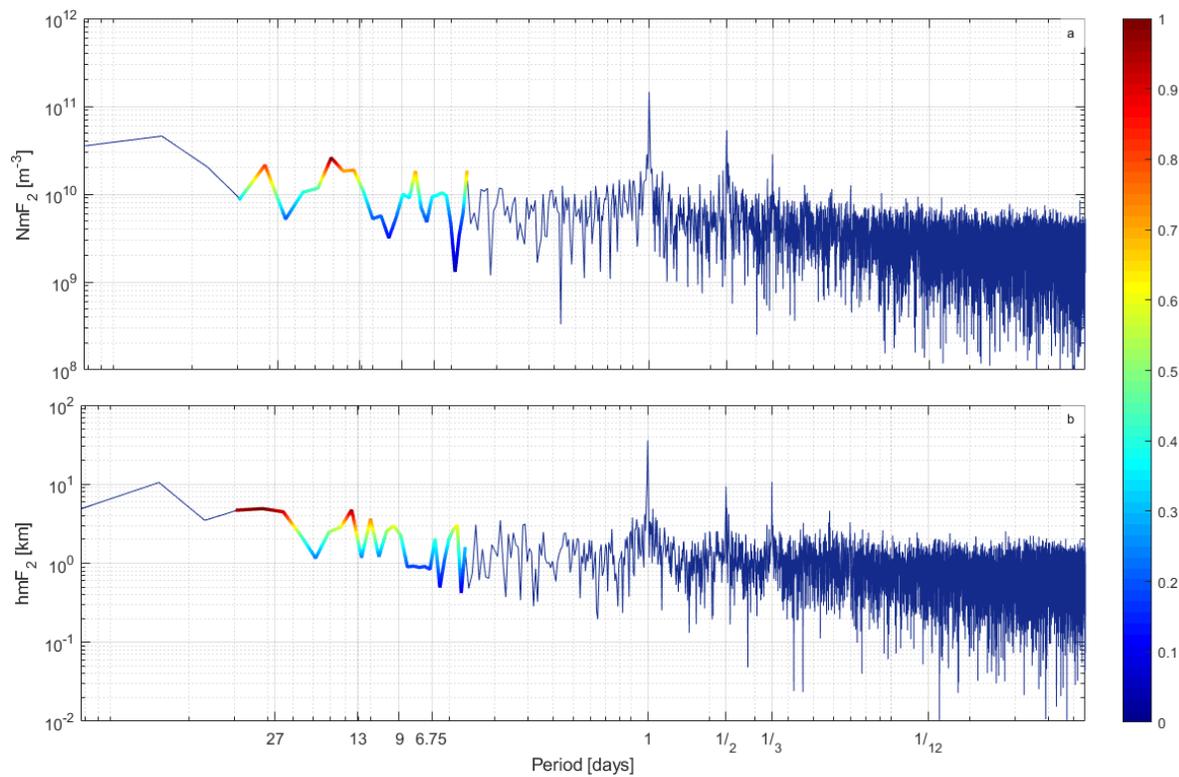

*Figure 4. Sample spectra for the Millstone Hill ionosonde, showing the a. $NmF_2$ and b. $hmF_2$ spectrum. The low-frequency portion of the spectrum is color coded according to a normalization performed on the period interval 5-36 days.*

Equivalent and related spectral features can be observed in the spectra characterizing the solar wind and geomagnetic indices, as can be seen in figure 5. The spectral features in this case are the direct results of the presence of periodic CIR/HSS structures, and their magnetospheric impact. Additionally, while the solar wind is a complex system, it is not subject to multiple types of forcings in the same manner as the thermosphere-ionosphere, allowing for the spectral peaks at $P_1$, $P_2$, $P_3$ and $P_4$ to have a higher signal-to-noise ratio and a lower spectral width.

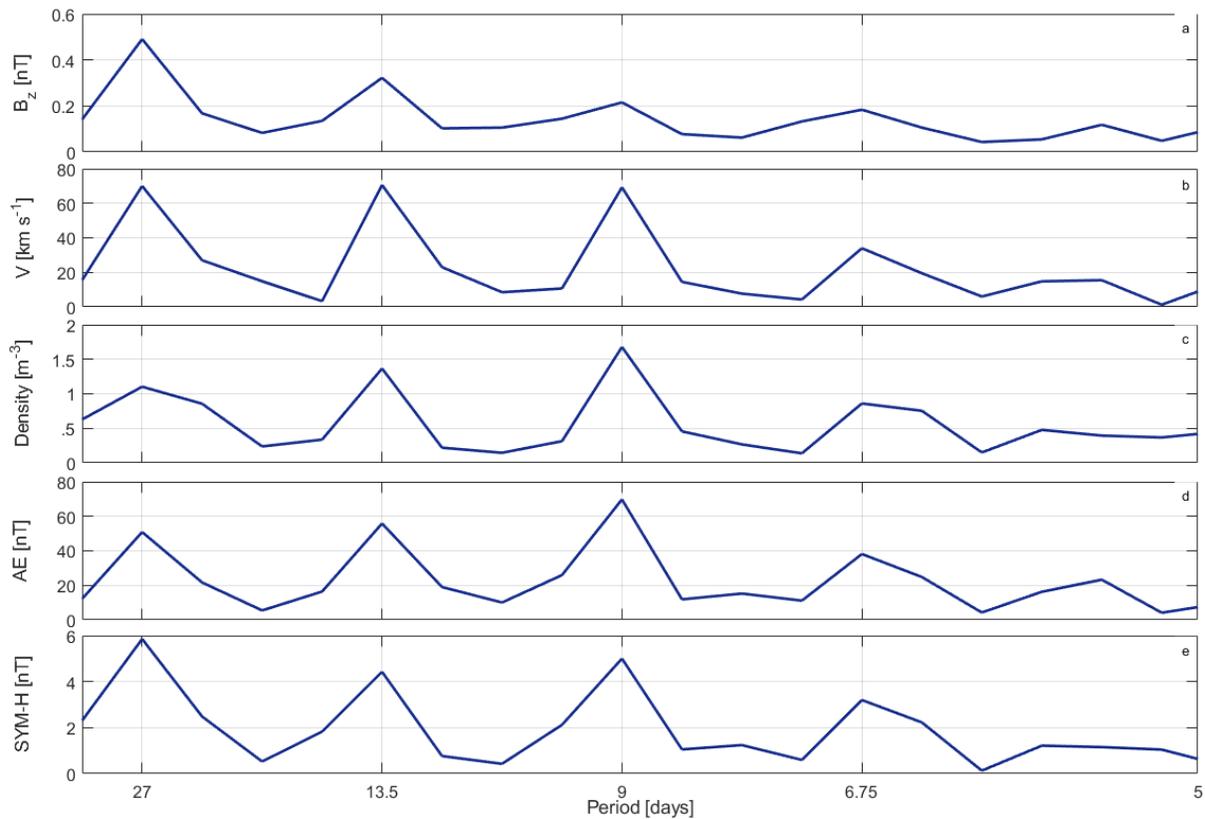

*Figure 5. Same as fig. 3, but for the solar wind parameters and geomagnetic indices.*

The causal link between periodic HSS/CIR events, associated geomagnetic storms (shown in fig. 5) and the periodic ionospheric response (shown in fig. 4) is well established. Characterized by periods $P_{1-4}$, it is superimposed onto the background variability of the thermosphere-ionosphere, created by solar irradiance, normal geomagnetic activity and GWs. A specific storm can be the main source of energy and momentum at some locations, for time intervals from several hours to at most a few days. Since the dataset we used has a length of 5 months, it is expected that, while significant, the periodicities induced by HSS/CIRs won't be dominant, and the specific response at a given ionosonde location may be affected by local phenomena. In order to highlight the global ionospheric response to HSS/CIRs while minimizing the impact of local conditions, the hmF$_2$ and NmF$_2$ spectra at each location are normalized to 1. Figure 4 exemplifies this procedure for a single station, highlighting the presence of spectral peaks while minimizing the impact of the spectral background, which may vary depending on location. Combining the results from all available stations in a mixed line-color plot yields a pseudo-2-D plot (figure 6), describing the ionospheric response as a function of period and latitude. Additionally, stations from specific latitude sectors are grouped together.

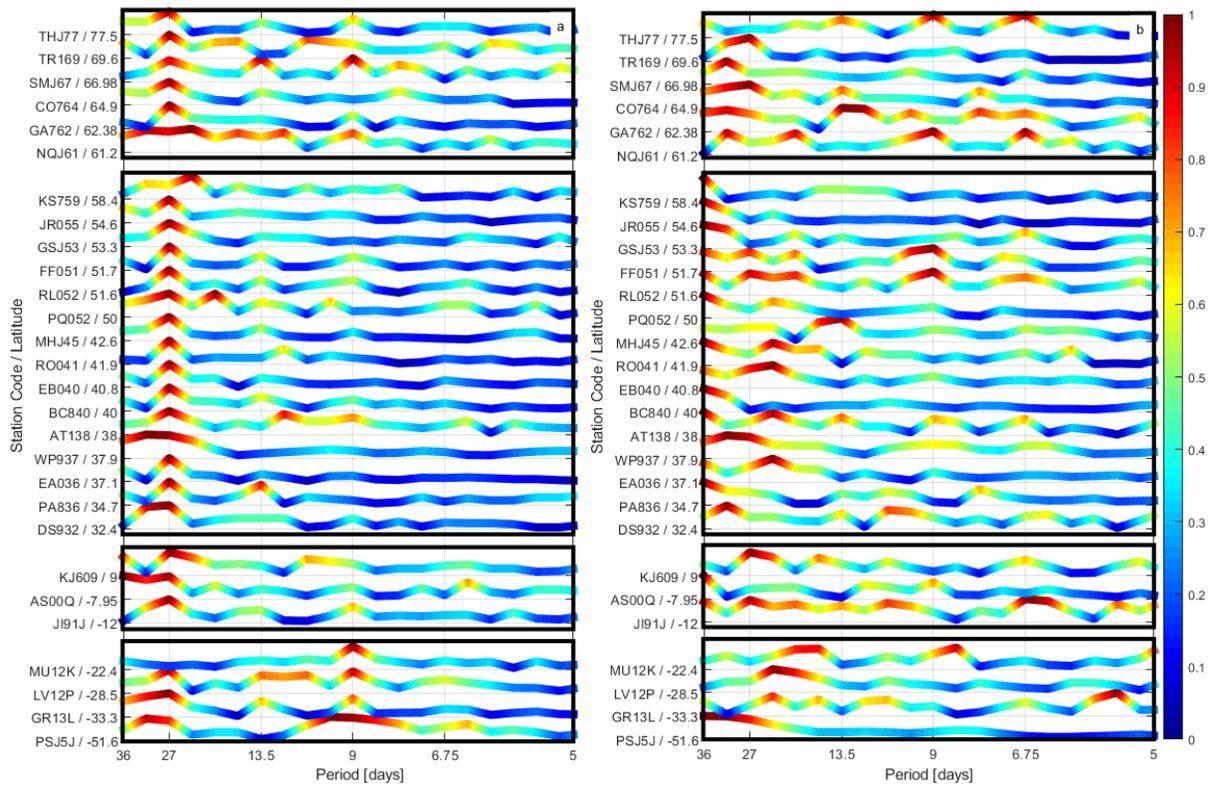

*Figure 6. Pseudo 2-D plot showing the low-frequency amplitude spectra as a function of latitude for a. hmF$_2$ and b. NmF$_2$. Each line represents the normalized amplitude spectrum for a different*

*ionosonde station, additionally color coded to highlight spectral features present at multiple locations. The stations are grouped into four categories: high latitude, northern hemisphere mid-latitude, low-latitude and southern hemisphere mid-latitude; with each group encased in a rectangle.*

3. Results and discussion

The hmF$_2$ and NmF$_2$ spectra depicted in figure 6 contain peaks causally linked to those observed in the solar wind and magnetospheric spectra in figure 5. The sequence of HSS/CIR events in the solar wind creates a magnetospheric response with the same dominant periodicities of 27, 13.5, 9 and 6.75 days, generated by the Suns rotation period, as well as the width and separation of coronal holes (*Lei et al.*, 2011). By contrast, the ionospheric response is characterized by a much more complex structure. While spectral peaks with periods P=P$_{1-4}$ are present at the majority of locations, secondary peaks are identified with intermediary periods, in the intervals 15-21, 10-12 and 7-8 days. Additional peaks are identified at smaller periods for a reduced number of locations, and will not be the focus of this study. The likely cause for of these additional periodicities is the non-linear nature of the energy and momentum transfer mechanisms associated with geomagnetic storms, manifested in the spectral domain by shifting a portion of the power contained in the solar wind and magnetospheric spectra to the secondary spectral peaks observed in the ionospheric spectra.

In the time domain, this may manifest in a number of ways:

i. the ionospheric response to specific storms may be different depending on geographic location, with differences in the start time and duration of the observed perturbations, as can be seen in figures 1 and 2;

ii.  at a given location the response to successive storms may exhibit different delays or durations, resulting in a slightly different spectral response;

Table 2. List of values for the hmF$_2$ and NmF$_2$ indices, I$_h$ and I$_N$, for each station, together with geographical coordinates.

| Station Code | Latitude (°) | Longitude (°) | I$_h$ | I$_N$ |
|---|---|---|---|---|
| THJ77 | 77.5 | 290.8 | 4 | 3 |
| TR169 | 69.6 | 19.2 | 3 | 2 |
| SMJ67 | 66.98 | 309.06 | 3 | 1 |
| CO764 | 64.9 | 212 | 2 | 3 |
| GA762 | 62.38 | 215 | 3 | 3 |
| NQJ61 | 61.2 | 314.6 | 2 | 2 |
| KS759 | 58.4 | 203.6 | 1 | 2 |
| JR055 | 54.6 | 13.4 | 1 | 1 |
| GSJ53 | 53.3 | 299.7 | 3 | 3 |
| FF051 | 51.7 | 358.2 | 4 | 2 |
| RL052 | 51.6 | 358.7 | 3 | 3 |
| PQ052 | 50 | 14.5 | 3 | 1 |
| MHJ45 | 42.6 | 288.5 | 3 | 3 |
| RO041 | 41.9 | 12.5 | 2 | 2 |
| EB040 | 40.8 | 0.5 | 3 | 2 |
| BC840 | 40 | 255 | 4 | 2 |
| AT138 | 38 | 23.5 | 3 | 3 |
| WP937 | 37.9 | 284.5 | 1 | 3 |
| EA036 | 37.1 | 353.3 | 3 | 2 |
| PA836 | 34.7 | 239.4 | 2 | 2 |
| DS932 | 32.4 | 260.2 | 4 | 2 |
| KJ609 | 9 | 167.2 | 3 | 2 |
| AS00Q | -7.95 | 345.6 | 3 | 1 |
| JI91J | -12 | 283.2 | 4 | 2 |
| MU12K | -22.4 | 30.9 | 3 | 2 |
| LV12P | -28.5 | 21.2 | 4 | 0 |
| GR13L | -33.3 | 26.5 | 4 | 2 |
| PSJ5J | -51.6 | 302.1 | 3 | 1 |

iii.  the hmF$_2$ storm response has a more pulse-like response, similar to the structures observed in the solar wind and the associated geomagnetic responses (figure 3), while in general the NmF2 response is more complex, with a positive and negative storm phase.

Thus, the spectral behavior of NmF2, namely its $P_{1-4}$ periodicities, is somewhat mingled compared to the more harmonic response of hmF2 which closely follows the recurrence of storms triggered by the sequence of HSS/CIR interplanetary events. This effect might have an additional impact on the global ionospheric response as the two main ionospheric parameters, the altitude of the F-layer and its maximum density do not react in phase with the external forcing.

The non-linear transfer is more pronounced for NmF2, resulting in discrepancies between the ionospheric response in NmF2 and hmF2. Most notably for $P=P_1$, 25 out of 28 locations show a spectral peak in hmF2, while an equivalent peak in NmF2 is only present at 11 locations. The overall hmF2 response at a given location can be quantified through the use of an index, $I_h$, with possible integer values between 0-4, which we define as the total number of spectral peaks clearly evidenced with periods equal to $P_{1-4}$ at each location, regardless of their relative or absolute amplitude. More exactly, for a given ionosonde, $I_h=\text{sum}(I_{hi})$, with $I_{hi}=1$ if a spectral peak is observed at $P_i$, or 0 otherwise. Equivalently for NmF2, we define the $I_N$ index. The values for each station are listed in table 2. Since $I_h > I_N$ for 22 locations, the non-linear nature of the solar wind-magnetosphere-ionosphere coupling has a much greater impact on the plasma density in the F-layer, rather than on the altitude of the layer itself. This result is observed particularly in the additional spectral peaks in figure 6b.

The mean value of $I_N$ has a significant latitude dependence, with $\overline{I_N}=2.3$ at high-latitude, 2.2 at mid-latitude in the northern hemisphere, and only 1.6 at low-latitudes. Additionally, a seasonal effect may also influence the ionospheric storm response, since $\overline{I_N}=1.25$ in the southern-hemisphere mid-latitudes. By contrast, $\overline{I_h}$=2.8 at high latitudes, increasing to 3.3 at low latitudes and 3.5 in the southern hemisphere. Additional investigation is required to clarify these effects, either using measurements from more locations, covering a wider time interval or involving additional data types. One interesting feature to note is that the values of $I_N$ are smaller than 4 at all locations, indicating that some or all of the dominant spectral peaks are shifted in the $NmF_2$ spectra, relative to the spectral peaks observed in the solar wind and geomagnetic indices.

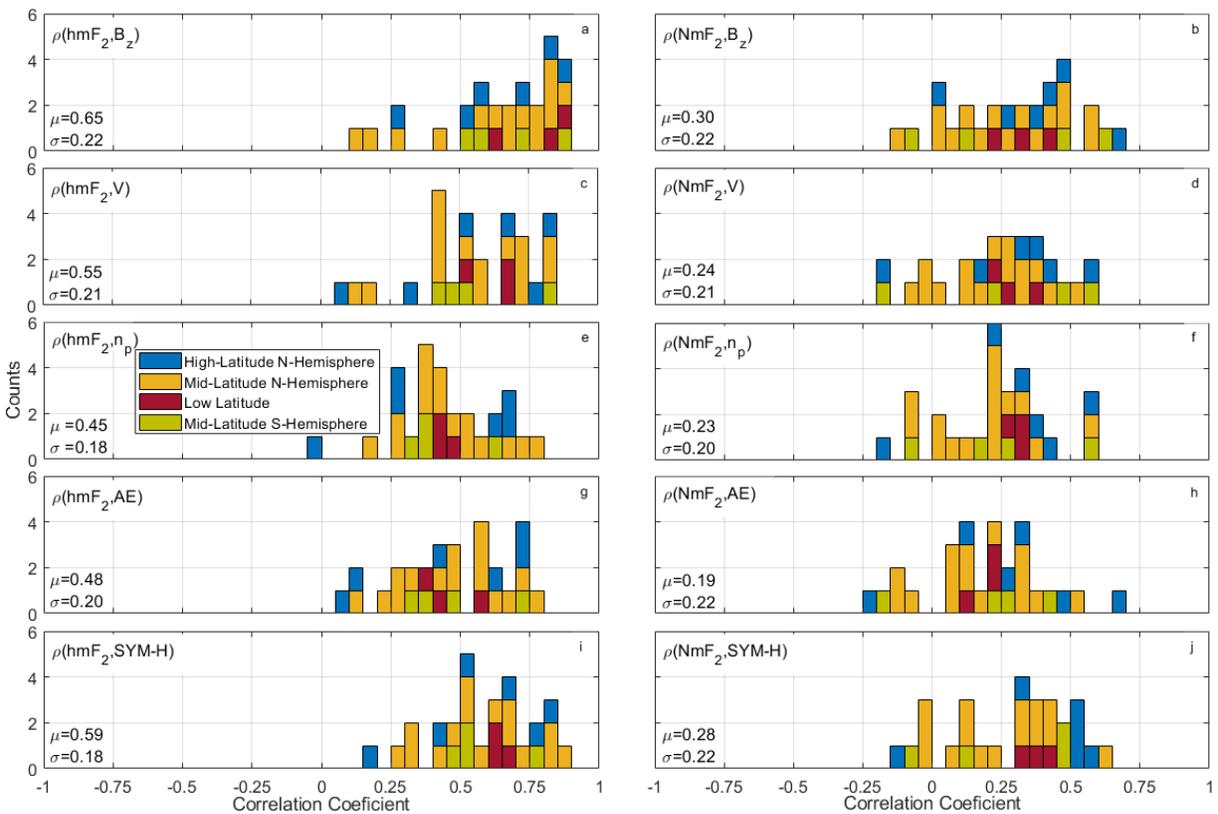

*Figure 7. Histograms of the correlation coefficients between the ionospheric parameter spectra and the spectra of the solar wind and geomagnetic indices: a. $hmF_2$ and $B_z$; b. $NmF_2$ and $B_z$; c. $hmF_2$ and V; d. $NmF_2$ and V; e. $hmF_2$ and $n_p$; f. $NmF_2$ and $n_p$; g. $hmF_2$ and AE; h. $NmF_2$ and AE; i. $hmF_2$ and SYM-H; j. $NmF_2$ and SYM-H. Each histogram covers the entire group of 28 ionosonde stations, while the color blocks depict the contributions from the subgroups containing the*

*locations within specific latitude ranges. For each pair of variables, the median and standard deviation is listed.*

To further investigate the correlation between the low-frequency response of the ionosphere to the recurrent sequence of interplanetary events, mediated by magnetospheric processes, we search for a quantitative measure of similarities between spectra of ionospheric parameters (hmF$_2$ and NmF$_2$) and spectra of solar wind (B, B$_z$ and V$_x$) and geomagnetic measures of activity (AE and SYM-H indices). The Pearson correlation coefficient is used:

$$\rho(A,B) = \frac{1}{M-1}\sum_{i=1}^{M}\left(\overline{\frac{A_i-\mu_A}{\sigma_A}}\right)\left(\frac{B_i-\mu_B}{\sigma_B}\right) \qquad (2)$$

where M is the total number of periods $P \in [5,36]$ days for which the spectral amplitude was determined, A and B are the two spectra for which the correlation coefficient is being determined, $\mu$ is the mean value and $\sigma$ is the standard deviation.

For each of the two ionospheric quantities (hmF$_2$ and NmF$_2$), and for each of the five solar wind and geomagnetic data types (B, B$_z$ V$_x$, AE and SYM-H), 28 correlation coefficients, $\rho(A_s, B)$ are determined, one for each of the ionosonde station locations, with a 95% probability that $\rho(A_s, B)$ highlights a real correlation if $\rho(A_s, B) > 0.22$ (*Pugh and Winslow*, 1966; *Bevington and Robinson*, 2003). The resulting histograms are used to estimate the probability distributions of the correlation coefficients, as shown in figure 7. Here, the global distribution describing all 28 locations is further detailed by highlighting the contributions of individual latitude sectors. Finally, he median for each distribution and associated standard deviation are also listed, under the assumption that a Gaussian is an accurate approximation of the real distribution. Note that in this case, unlike for $I_h$ and $I_N$, the whole low-frequency spectrum (periods between 36 and 5 days) was used, not just the spectral peaks. For both NmF$_2$ and hmF$_2$ the median correlation value is positive with all OMNI parameters, indicative of the solar wind-magnetosphere-ionosphere coupling. However, the median values for all distributions describing hmF$_2$ were higher than those of the equivalent NmF$_2$ distributions. The hmF$_2$ data exhibit a median correlation between 0.45 to 0.65 with the solar wind and geomagnetic parameters, while the range for the NmF$_2$ data is only between 0.19 to 0.30. Since the associated standard deviations are similar, 0.18-0.22 for hmF$_2$ and 0.20-0.22 for NmF$_2$, it can be concluded that the hmF$_2$ more closely follows the geomagnetic forcing. This is likely attributed to the more linear nature of the hmF$_2$ response, for which the positions of the spectral peaks align with those in the solar wind and geomagnetic indices. By contrast, in the NmF$_2$ spectra, the position of all or some of the dominant spectral peaks is shifted, lowering the correlation coefficients.

Both the hmF$_2$ and NmF$_2$ spectra exhibit the strongest correlation with solar wind B$_z$, and the least strong with n$_p$. Interestingly, of the two geomagnetic indices considered, a stronger correlation is observed with the spectral characteristics of SYM-H, suggesting that the ionospheric storm response can more strongly be attributed to penetrating electric fields in this case, rather than high-latitude Joule heating and particle precipitation. Similar conclusions have been previously reported by *Verkhoglya et al.* (2013) and *Vijaya Lekshmi et al.* (2008) using different approaches and investigating different periods of time. Penetrating electric fields are sudden events, closely coupled to the external driver. Therefore, their ionospheric effects might be characterized by a closer similarity, in terms of periodicities, with the driver spectral features. On the other hand, Joule heating at higher latitudes necessarily adds a time scale for the ionospheric response to the driver, linked to the particle heating itself and also to the time required for the associated ionospheric perturbation to propagate at lower latitudes. Finally, note that the subgroups of locations belonging to any given latitude interval do not display a noticeably different probability distribution function than that of the whole group of stations. This points to a truly global nature of the ionospheric storm response, although further investigation is required to determine to what extent this is valid for less impactful series of geomagnetic storms.

4. Summary and conclusions

In this study, we used ionosonde $hmF_2$ and $NmF_2$ data (figures 1 and 2) from 28 stations listed in table 1, as well as solar wind measurements and geomagnetic indices taken from the OMNI dataset (figure 3) to investigate the ionospheric impact of a sequence of ten HSS/CIR events during 2007-2008. Spectral analysis is used to highlight the low-frequency impact of this phenomena in the solar wind and geomagnetic activity (figure 5) and the associated ionospheric impact (figure 4). In order to account for the different spectral background at each location, the low-frequency range (periods between 5-36 days) of each spectrum is normalized to 1 (figure 4). While the solar wind and geomagnetic spectra contain local maxima at precisely $P_1=27$, $P_2=13.5$ $P_3=9$ and $P_4=6.75$ days (figure 5), the ionospheric spectra are characterized by a more complex structure, with secondary peaks adding to or replacing the main ones, particularly for $NmF_2$ (figure 6).

This increase in complexity is likely caused by the non-linear nature of the magnetosphere-ionosphere coupling. By analyzing both the $NmF_2$ and $hmF_2$ spectra, we show that this non-linearity is much more pronounced for $NmF_2$, and potentially latitude dependent. Two indices are introduced, $I_N$ and $I_h$, counting how many peaks with periods $P_1$, $P_2$, $P_3$ and $P_4$ occur at each location in the $NmF_2$ ($I_N$) and $hmF_2$ spectra, respectively. The results listed in table 2 show that $I_N<I_h$ in the majority of cases, with $I_N<4$ at all locations, meaning that no location shows all the four periodicities observed for the forcing. Additionally, the non-linearity of the ionospheric response in NmF$_2$ may be latitude and seasonally dependent, although future work is necessary to clarify this, involving additional measurements and data types.

The coupling of the ionospheric parameters and the solar wind data (B$_z$, V and n$_p$) and geomagnetic indices (AE and SYM-H) is investigated quantitatively through estimating the degree of similarity between the spectral features of ionospheric height and density of the F-layer (through hmF$_2$ and NmF$_2$) and forcing (through solar wind variables Bz, V, n$_p$, and geomagnetic indices, AE and SYM-H). Such a quantitative measure was derived from a computation of the correlation coefficients between the spectra themselves. For each variable pair, a probability distribution function describing the level of spectral correlation is constructed using spectra from all 28 ionosondes (figure 7). The resulting distributions are characterized by higher median values for hmF$_2$, whose spectra closely follows those characterizing the sequence of HSS/CIR interplanetary events. At the same time the NmF$_2$ median correlation is approximately 0.3 lower due to its more complex spectra. This indicates that the altitude of the F2-layer and its maximum density exhibit a fundamentally different response to the external forcing.

Since in general the storm response of NmF$_2$ is stronger than that of hmF$_2$, this result is further evidence of the non-linear nature of the solar wind-magnetosphere-ionosphere coupling: the local maxima in the NmF$_2$ spectra are not located at P$_1$-P$_4$, or not only at these periods, decreasing the correlation. Finally, the ionospheric spectra is more strongly correlated with the solar wind B$_z$ than V or the n$_p$, and also a stronger correlation with SYM-H than AE. This suggests that for the low-frequency ionospheric response, penetration electric fields play a more important role than the high-latitude Joule heating.

Acknowledgements

This work is part of a project that has received funding from the European Union's Horizon 2020 research and innovation programme under the Marie Skłodowska-Curie grant agreement No. 800215, and also by the Romanian Ministry of Research through UEFISCDI PCCDI grant VESS (contract no. 18/2018) and National Core Program (LAPLAS 2019). The authors wish to thank T. W. Bullett and NOAA-NCEI for providing access to the ionosonde data necessary for this work, which is accessible at http://ngdc.noaa.gov/ionosonde/data. Finally, we would like to thank the two reviewers for their suggestions and comments that helped improve the quality of our paper.